\documentclass[english, aps, prl, longbibliography, preprint]{revtex4-1}
\usepackage[T1]{fontenc}
\usepackage[latin9]{inputenc}
\usepackage{color}
\usepackage{amssymb}
\usepackage{graphicx}
\usepackage{amsmath}
\usepackage{breakurl}
\usepackage{graphicx}
\usepackage{verbatim}
\usepackage[unicode=true,pdfusetitle, bookmarks=true,bookmarksnumbered=false,bookmarksopen=false, breaklinks=false,pdfborder={0 0 1},backref=false,colorlinks=true]{hyperref}
\hypersetup{linkcolor=blue, citecolor=blue, urlcolor  = blue}

\newcommand\citepar\citep
\newcommand\citebyname\citet
\newcommand\micron{{\rm \mu m}}

\newcommand\Reynolds{{\it Re}}


\begin{document}

\title{Separated Rows structure of vortex streets behind triangular objects}
\author{Ildoo Kim}
\email{ildoo.kim.phys@gmail.com}
\affiliation{School of Engineering, Brown University, Providence, Rhode Island 02912}
\date{\today}

\begin{abstract}
We discuss two distinct spatial structures of vortex streets.
The `conventional mushroom' structure is commonly discuss in many experimental studies, but the exotic `separated rows' structure is characterized by a thin irrotational fluid between two rows of vortices.
In a two-dimensional soap film channel, we generated the exotic vortex arrangement by using triangular objects.
This setup allows us to vary the thickness of boundary layers and their separation distance independently.
We find that the separated rows structure appears only when the boundary layer thickness is less than 40\% of the separation distance.
We also discuss two physical mechanisms of the breakdown of vortex structures. 
The conventional mushroom structure decays due to the action of viscosity, and the separated rows structure decays because its arrangement is hydrodynamically unstable.
\end{abstract}

\maketitle

\section{Introduction} 

When a stream of flow interacts with a fixed boundary, vorticity is created in the boundary layer and discharged into the fluid to form a self-organized pattern known as von K\'{a}rm\'{a}n vortex street.
Previous studies on this celebrated topic include the formation \citepar{Williamson:1996tn}, Strouhal-Reynolds number relations \citepar{Williamson:1998ti, Fey:1998vl, Roushan:2005un
}, control \citepar{choi:2008bc}, etc.

\begin{figure}
\begin{centering}
\includegraphics[width=8cm]{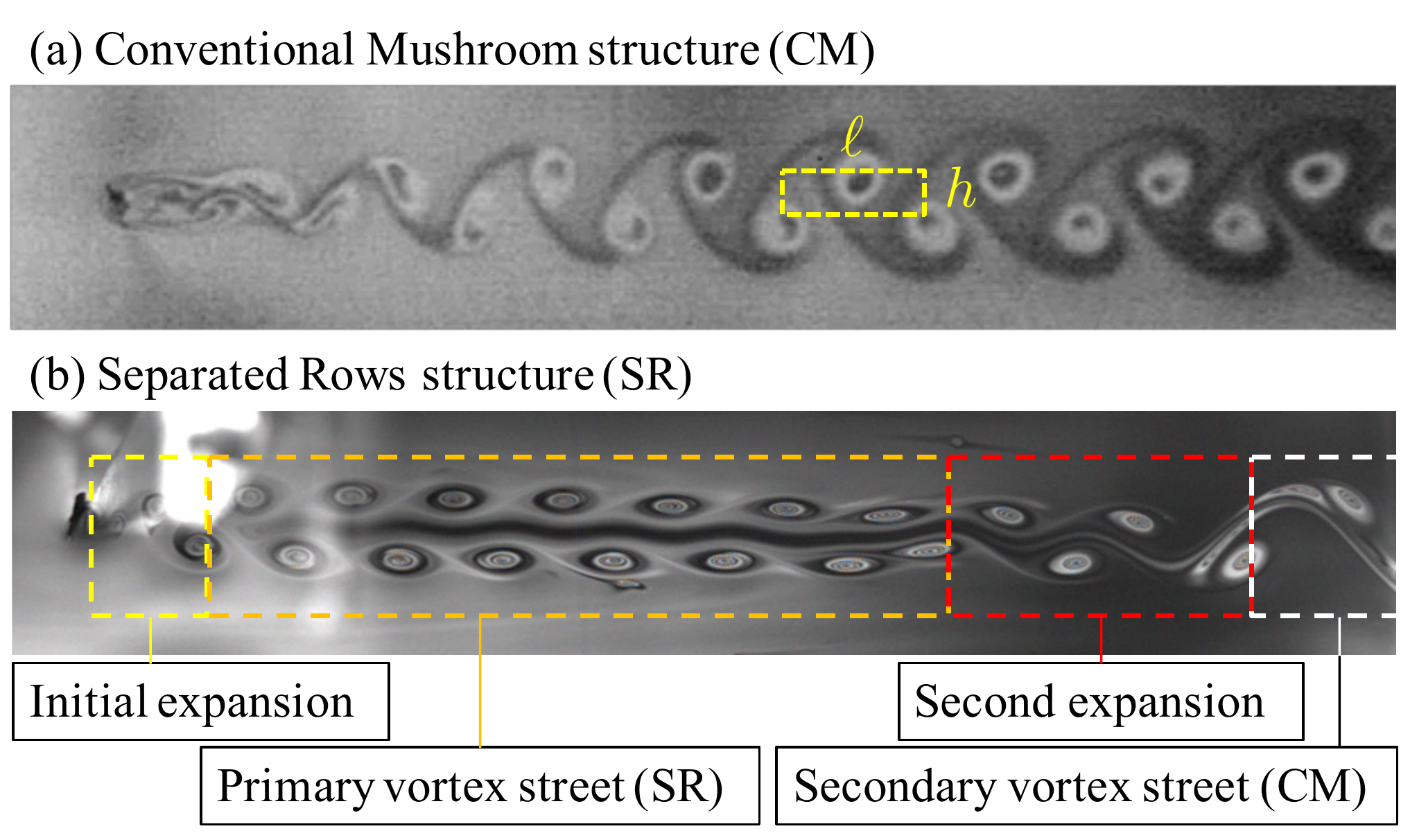}
\par
\end{centering}
\caption{
Two wake structures in a flowing soap film. 
(a) The Conventional Mushroom structure (CM) is created using a circular object.
The K\'{a}rm\'{a}n ratio is measured to be 0.27. 
(b) The Separated Rows structure (SR) is created using a triangular object. 
This exotic structure is characterized by a thin irrotational fluid between two vortex rows and the higher $r_K$, which is measured to be 0.45.
The Reynolds number is same ($\Reynolds=660$) in both cases.
\label{fig:type2sample}}
\end{figure}

We report vortex streets whose spatial structure is different from commonly observed ones, as shown in Fig. \ref{fig:type2sample}.
These two spatial structures are morphologically different as follow.
First, two rows of the vortices in (b) are separated by a thin irrotational fluid while the vortices in (a) are intermingled.
The structure in (b) resembles the Kelvin-Stuart cat's eye flow, which is a quintessential of the shear layer instability.
The observation suggests that there are two length scales in the problem, where one is the thickness of the boundary layers and the other is the separation distance between them.
Second, the K\'{a}rm\'{a}n ratios is higher in (b) than in (a).
The K\'{a}rm\'{a}n ratio $r_K$ is defined as the ratio between $h$, the distance between two rows of vortices, and $\ell$, the distance between two vortices in the same row, i.e. $r_K\equiv h/\ell$, and has an implication regarding the stability of the vortex street.
Using the point vortex model, \citebyname{vonKarman:1911vi} showed that only a vortex arrangement having $r_K=\cosh^{-1}(\sqrt{2})/\pi\approx0.281$ is stable.
His prediction of the narrow range of the stability was at odds with the experiments which showed that vortex streets exist over a range $0.25<r_{K}<0.52$ \citepar{Goldstein-38}, and \citebyname{Hooker:1936tz} showed that the action of the viscosity broadens the range of the stability.

The wake structure similar to Fig. \ref{fig:type2sample}(b) has been perceived to a precursor of the breakdown of a primary vortex street \citepar{Taneda:1959wj,Henderson:1996tc,Inoue:1999wi,Vorobieff_2002,Wang_2010,Kumar_2012,Dynnikova:2016is}, but we find that this is not always the case.
It had been considered as an interim structure that occurs only when the primary vortex street, which is an archetypal structure in Fig. \ref{fig:type2sample}(a), breaks down and is transformed to the secondary vortex street, which is also an archetypal structure.
However, we find that this exotic structure also occurs in the primary wake without the presence of the archetypal structure when we use triangular objects to generate vortex streets.
This observation indicates that the spatial structure in (b) is neither a far field phenomenon nor a component of the wake transformation.
For the clarity of the ensuing discussions, we name the archetypal structure in (a) as the Conventional Mushroom structure (CM) and the ones in (b) as the Separated Rows structure (SR). 

In this paper, we investigate the role of the boundary layer to the formation and structure of vortex streets.
Previous observations of vortex streets indicate that the spatial structure depends on the initial condition that produces the vorticity \citepar{Kim:2015jp}.
However such an effect has not been studied in a systematical manner because the circular cylinder, which is being used in most experiments in this area, has only a single length scale (the diameter) that determines the downstream flow structure.
The current study uses bluff bodies with the triangular cross section, and the height $H$ and the base $D$ of the triangle are independently varied.
By using the triangular object of various aspect ratios $r_a=H/D$, we independently vary the boundary layer thickness $\delta$ and the separation distance $D$ between two layers.
This setting allows us to create different vortex structures and to manipulate the transition between them.

We find that there is a critical ratio between $\delta$ and $D$ that determines the stability of vortex streets.
When $\delta/D>0.4$, the vortex street is created in CM and is stable over the range of our experiments.
When $\delta/D<0.4$, the vortex street is created either in CM or in SR, and the vortex street is unstable and decays downstream.
If the vortex street was created in CM, it gradually broadens to transform to SR.
If the vortex street was created in SR, or has transformed to SR, it transforms back to SR through a sudden and full-scale rearrangement of vortices, as previously known as the secondary instability \citepar{Williamson1993,Cimbala1988,Inoue:1999wi,Vorobieff_2002,Wang_2010,Kumar_2012,Dynnikova:2016is}. 

It is inferred from our results that the stability of vortex streets are determined by two mechanisms.
In CM, the vortex street decays because the vorticity is dissipated through mixing of two rows.
In SR, however, the dissipation through mixing does not happen, but the vortex street decays because of its unstable spatial arrangement.
These two mechanisms act interchangeably, and the alternating sequence of CM and SR may be observed in a long channel.

\section{Experimental setup}

The experiments are carried out using an inclined soap film channel \citepar{Kim:2015jp,Georgiev:2002kg}. 
Our channel is approximately 200 cm long and 5 cm wide, and we use the soap solution made from 2\% commercial dish soap (P\&G Dawn), 5\% glycerol, and rest distilled water. 
The kinematic viscosity of the solution $\nu\simeq0.013\pm0.001\,\rm cm^2/s$, measured by a Cannon viscometer. 
The flow speed and the thickness of the channel can be adjusted by the flow rate. 
Throughout the study, we fix the flow rate at 0.09 $\rm cm^3/s$.
Under this condition, the flow speed $U$ is $60\pm3$ cm/s, and the thickness of the film is approximately $3\,\micron$.
The film is only slightly compressible because the flow speed is only 16\% of the Marangoni wave speed \citepar{Kim:2017dn}, indicating that the fluid medium is quasi two-dimensional.

\begin{figure}
\begin{centering}
\includegraphics[width=7cm]{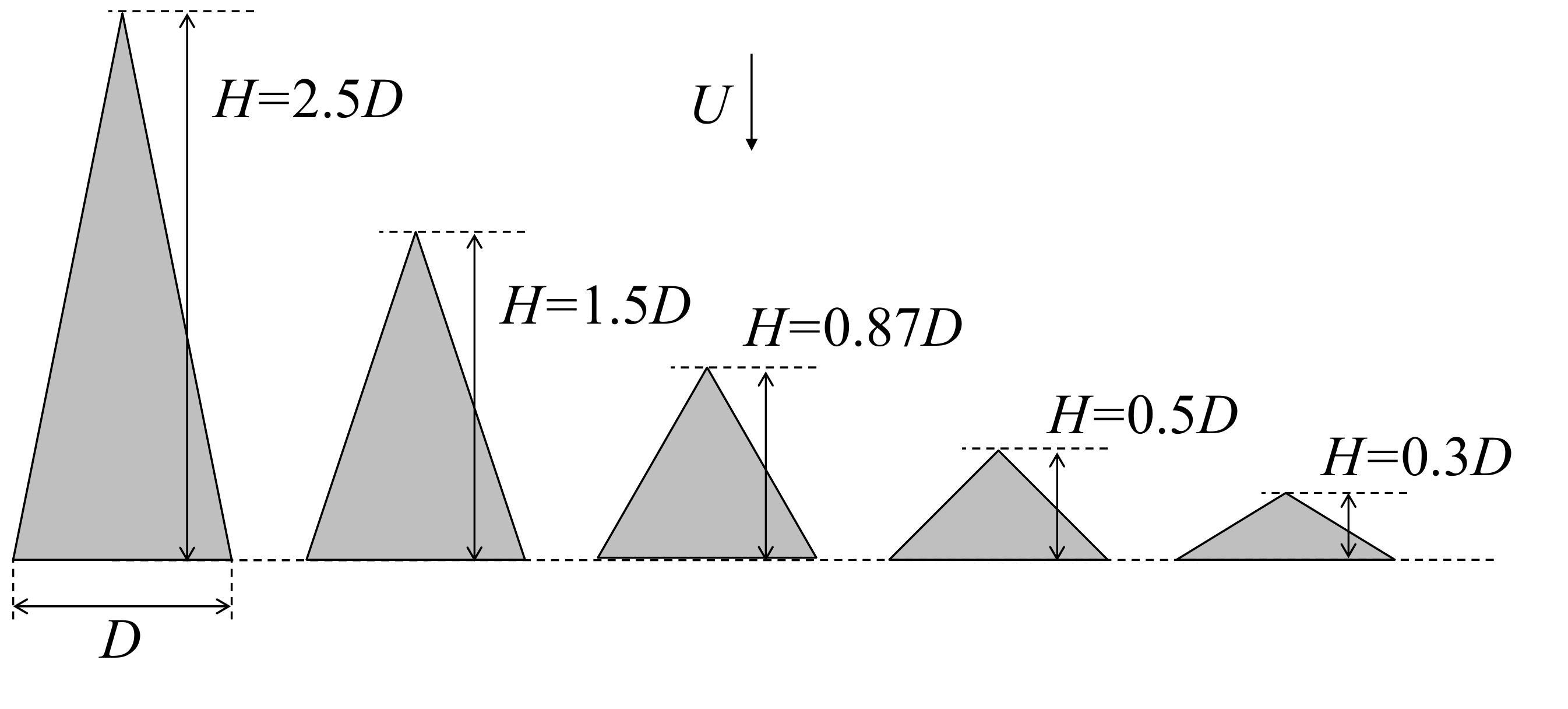}
\par
\end{centering}
\caption{
Cross-sections of tapered rods used in the experiments. 
These rods are characterized by the aspect ratio $r_a\equiv H/D$, where $H$ and $D$ are the height the base.
Vortex streets are generated by inserting a tapered rod to the soap film, as the apex of the triangle faces the flow.
\label{fig:experimentalsetup}}
\end{figure}

We generate vortex streets by inserting tapered rods to the soap film.
These rods have triangular cross-sections each at five different aspect ratios $r_a\equiv H/D=$0.3, 0.5, 0.87, 1.5 and 2.5, where $H$ is the height and $D$ is the base of the triangle, as depicted in Fig. \ref{fig:experimentalsetup}.
The apex of the triangle faces the flow.
The rods are made of titanium, and each rod is carefully tapered to maintain the unique $r_a$ to the tip size 50 $\micron$.
Therefore, depending on the insertion depth of the rod to the soap film, $D$ is adjusted from 0.5 cm to $50\,\micron$.
We define the Reynolds number using the base of the triangle, i.e. $\Reynolds\equiv UD/\nu$.
Since $U$ and $\nu$ are fixed in our setup, $\Reynolds$ is solely determined by $D$.

Using these rods of various aspect ratios, we investigate the effect of the boundary layer thickness on the structure and the stability of vortex streets. 
The boundary layer thickness $\delta$ is thicker when the aspect ratio of the rod is higher because the flow and the boundary interact for a longer time.
We use the empirical relation
\begin{equation}
\delta\simeq6\left(r_{a}^{2}+0.25\right)^{1/4}Re^{-1/2}D,
\label{eq:boundary-layer-thickness}
\end{equation}
whose derivation is presented in the Appendix.

Two video cameras are used for our measurements. 
A normal-speed camera (30 fps) is attached to a long-range microscope to monitor the insertion depth of the rods to the soap film, and $D$ is precisely determined.
The vortex streets in the soap film is illuminated by a monochromatic light (low-pressure sodium lamp), and the interferogram is captured by a high-speed camera (Phantom V5, Vision Research) at up to 1000 fps.
The visualization technique allows us to directly measure the flow structure.
The field of view of the high-speed camera is approximately 8 cm $\times$ 8 cm.
It is wide enough to cover $40D$ to $100D$ depends on $\Reynolds$, and the flow structures are monitored far downstream.

\section{Results and Discussions}

\subsection{Wake structures}

\begin{figure*}
\begin{centering}
\includegraphics[width=\textwidth]{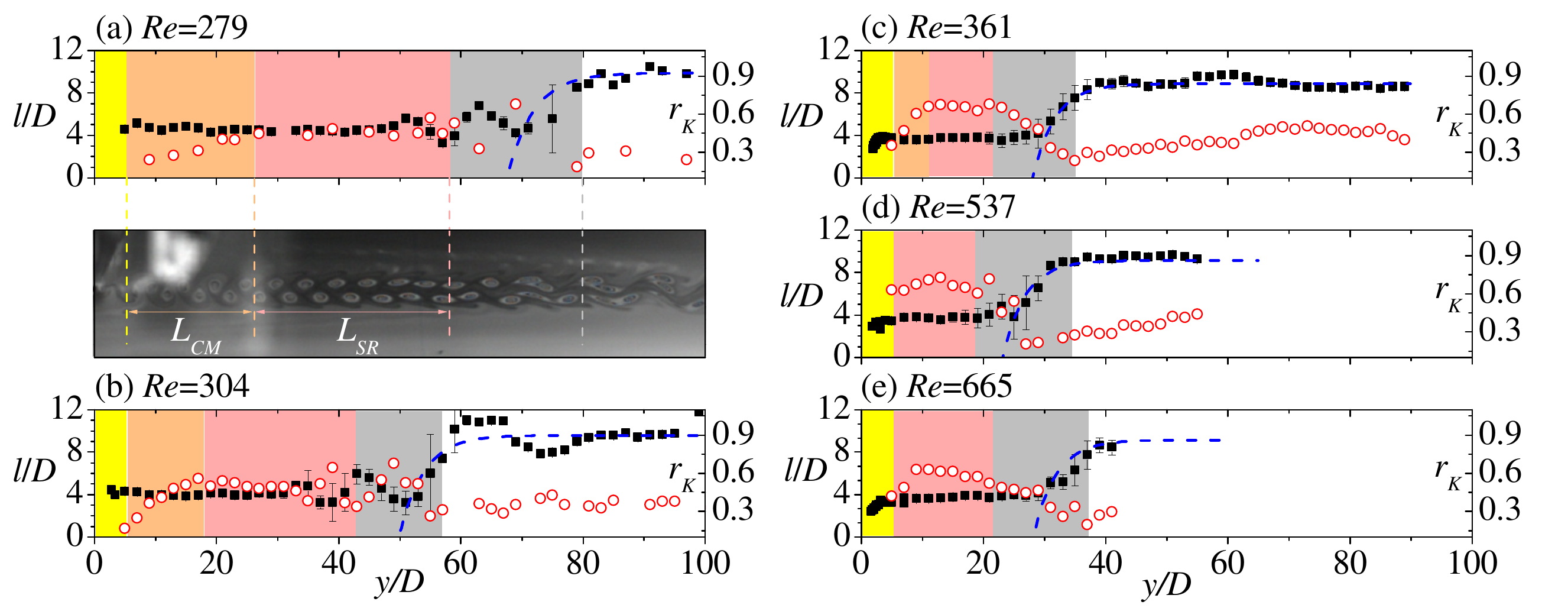}
\par
\end{centering}
\caption{
The characterization of vortex structures.
The longitudinal spacing between vortices $\ell/D$ (closed squares) and the K\'{a}rm\'{a}n ratio $r_{K}$ (open circles) are measured with respect to the downstream distance $y/D$ for vortex streets behind the tapered triangular rod of $r_{a}=0.5$. 
A vortex street is separated into following stations: (i) the expansion zone (shaded yellow), (ii) primary wake in CM (shaded orange) or in SR (shaded pink), (iii) the second expansion zone (SR$\rightarrow$CM transition zone, shaded grey), and (iv) the secondary wake zone.
We notice that $\ell$ rapidly increases in the expansion zones and is stable and unchanged in the wake zones, and its final approach is characterized as the exponential expansion in Eq. (\ref{eq:exponential_expansion}), as indicated by blue dotted lines. 
Unlike $\ell$, K\'{a}rm\'{a}n ratio $r_{K}$ is gradually increasing when the vortex street is in CM.
\label{fig:transient}}
\end{figure*}

The wake structure is clearly characterized by measuring $\ell$ and $r_K$, the distance between two vortices in the same row and the K\'{a}rm\'{a}n ratio, respectively.
In Fig. \ref{fig:transient}, we present our measurements of $\ell$ and $r_K$ for vortex streets generated by using the rod of $r_a=0.5$, at five different $\Reynolds=$279, 304, 361, 537, and 665 ($D$=0.047, 0.053, 0.062, 0.095 and 0.127 cm).
The measurements are plotted with respect to the downstream distance $y$ and normalized by the size of the rod $D$.
We find that a vortex street is separated into the following stations: (i) the initial expansion zone, (ii) the primary wake zone, (iii) the second expansion zone, and (iv) the secondary wake zone.

In the initial expansion zone, the vorticity created in the boundary layer is discharged into the fluid and organized into the primary vortex street.
In Fig. \ref{fig:transient}, this region is shaded yellow.
The initial development of the spatial pattern has been discussed by \citebyname{Kim:2015jp}.
They have shown that the initial expansion is characterized by the exponential function $\ell =\ell_1[1-\exp(-y/y_1)]$, where $\ell_1$ is an asymptotic value and a relaxation length $y_1\sim D$.
Our experiments show a similar trend in the initial expansion, and the extent of the initial expansion zone is less than $5D$ from the rod.

In the primary wake zone, the vortex street is either in CM followed by SR as displayed in Fig. \ref{fig:transient}(a-c) or entirely in SR as in (d-e).
The CM part is shaded in orange, and the SR part is shaded in pink.
For the primary wake, $\ell$ is stable and does not change over $y$, regardless of the wake structure.
Our measurement shows that $\ell\simeq 4D$, and this is consistent with the previous reports using circular and squares rods \citep{Roushan:2005un,Kim:2015jp}.
Unlike $\ell$, $r_K$ depends on the wake structure.
When the vortex street is in CM, $r_K$ gradually increases with $y$ due to a subtle rearrangement of vortices that causes the vortex street to broaden.
The rate of the broadening increases as $\Reynolds$ increases, and the broadening continues until the wake structure is transformed to SR.
When the vortex street is SR, for either (a-c) or (d-e) scenario, both $h$ and $\ell$ do not depend on $y$.
We find that $r_K$ of SR ranges from 0.45 to 0.8 depending $\Reynolds$.
This range of $r_K$ is substantially higher than the theoretical prediction from the point vortex model $r_K=\cosh^{-1}(\sqrt{2})/\pi\thickapprox0.281$.
These SR vortex streets with high $r_K$ are unstable and break down downstream, followed by a violent rearrangement of vortices in the secondary expansion zone.

The broadening of vortex street in CM is caused by the decay of vorticity through the mixing of two vortex rows.
\citebyname{Birkhoff1953} has shown that in any plane flow satisfying the Navier-Stokes equation, the moment of the vorticity is constant,
which infers the conservation of $h\kappa$, where $\kappa$ is the vorticity of vortices.
In the primary wake zone of our experiments, two vortex rows of CM are close enough to get mixed.
Therefore the strength of vortices $\kappa$ diminishes, which causes $h$ to increase.
On the other hand, the two vortex rows in SR are well-separated by the irrotational fluid between them.
In this case, $\kappa$ remains unchanged, and the broadening does not take place.

We calculate the broadening of the vortex street by using a simple model.
The linear model assumes that the decay rate of $\kappa$ is proportional to $\delta/h$.
Simply,
\begin{equation}
U\frac{d\kappa}{dy}=-\frac{\kappa\delta}{h\tau},
\end{equation}
where $\tau$ is a time scale of the fluid medium that is supposedly independent of $\kappa$ and $D$.
Using the conservation law $h\kappa=h_0\kappa_0$, where $h_0$ and $\kappa_0$ are values at $y=y_0$ where the primary wake zone begins, the solution of the model predicts that $h$ would linearly increase with $y$.
The K\'{a}rm\'{a}n ratio is derived as
\begin{equation}
r_K=r_K(y_0)+\frac{(y-y_0)}{D}\frac{\delta }{U\tau(\ell/ D)}
\label{eq:r_K_broadening}
\end{equation}
Considering that $\delta\propto D^{1/2}$ and $\ell\propto D$, we infer from Eq. (\ref{eq:r_K_broadening}) that the rate of broadening increases with $\Reynolds$ as we observed in the experiments.

\begin{figure*}
\begin{centering}
\includegraphics[width=\textwidth]{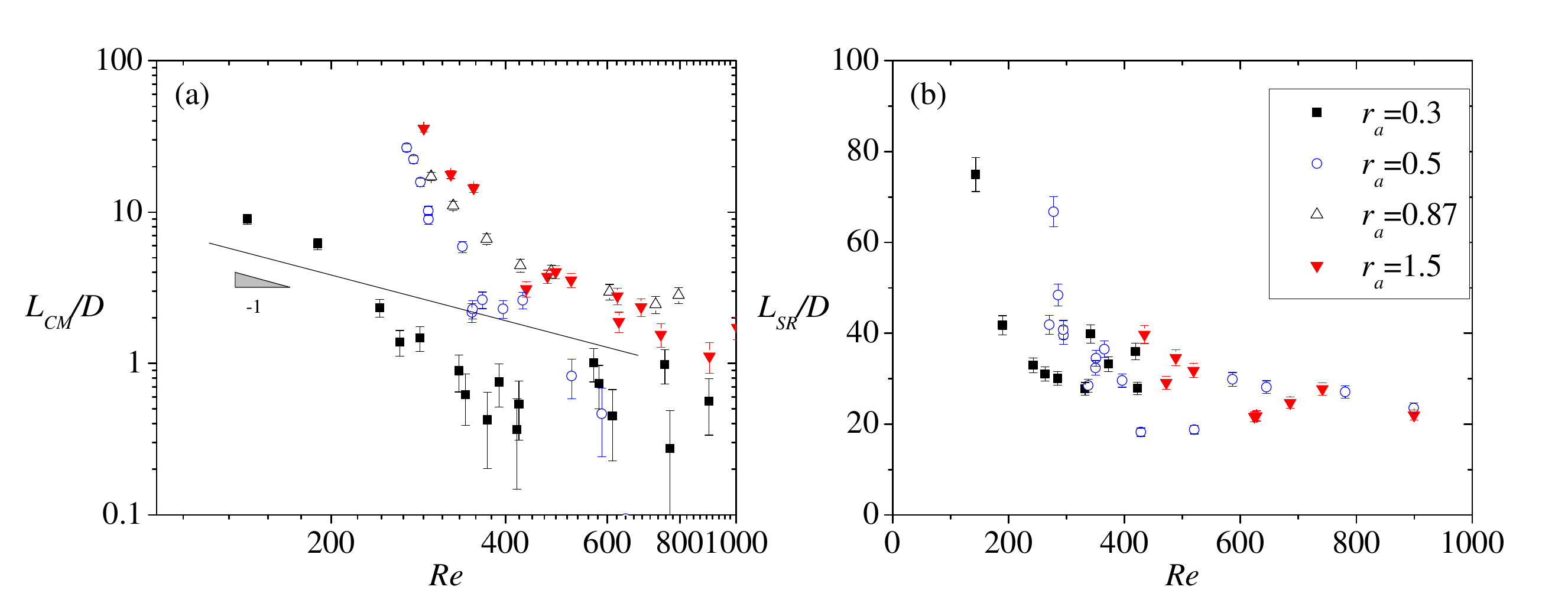}
\par
\end{centering}
\caption{
The spatial extents of vortex structures.
The length of (a) CM part and (b) SR part in the primary wake zone are plotted with respect to $\Reynolds$. 
Both $L_{CM}$ and $L_{SR}$ approach asymptotic values, 0 and $\sim 20D$ respectively, as $\Reynolds$ increases, but they increase as $\Reynolds$ decreases.
Also, a rough scaling relation $L_{CM}\propto \Reynolds^{-1}$ is observed.
\label{fig:spatial_extent}}
\end{figure*}

Vortex structures in Fig. \ref{fig:transient} are unstable and decays downstream.
We quantify the instability by measuring their spatial extents $L_{CM}$ and $L_{SR}$, which are the length of CM and SR, respectively, in the primary wake zone.
In Fig. \ref{fig:spatial_extent}(a), we show that $L_{CM}/D$ approaches zero as $\Reynolds$ increases, indicating that the primary wake is purely SR without the appearance of CM.
As seen in (b), $L_{SR}$ approaches an asymptotic value $\sim 20D$ when $\Reynolds$ is high, suggesting that a finite time is required for the breakdown of SR vortex street.
We find that both $L_{CM}$ and $L_{SR}$ increase as $\Reynolds$ decreases, indicating that both structures are more stable when $\Reynolds$ is small.

We suspect that $L_{CM}$ is longer when $\Reynolds$ is small because the vortex broadening rate is smaller.
Using Eq. (\ref{eq:r_K_broadening}), we write
\begin{equation}
\frac{L_{CM}}{D}=(r_K^{(max)}-r_K^{(min)})\frac{U\tau\ell}{\delta D}
\label{eq:L_CM_1})
\end{equation}
where $r_K^{(min)}=r_K(y=y_0)$ and $r_K^{(max)}=r_K(y=y_0+L_{CM})$.
Using Eq. (\ref{eq:boundary-layer-thickness}) and $\ell=\ell_0+\alpha D$, where $\ell_0$ is a non-zero intercept and $\alpha$ is a proportionality constant \citepar{Kim:2015jp}, it becomes
\begin{equation}
\frac{L_{CM}}{D}=\frac{(r_K^{(max)}-r_K^{(min)})}{6(r_a^2+0.25)^{1/4}} \frac{U^2\tau\alpha}{ \nu}  \frac{1}{\sqrt\Reynolds} (1+\frac{U\ell_0}{\alpha\nu}\frac{1}{\Reynolds}).
\label{eq:L_CM}
\end{equation}
Our measurements suggests that $L_{CM}\propto \Reynolds^{-1}$.
Considering that two scaling relations are interlaced around $\Reynolds\simeq U\ell_0/\alpha\nu \approx 200$, Eq. (\ref{eq:L_CM}) is roughly consistent with the measurements.

In the second expansion zone (shaded grey in Fig. \ref{fig:transient}), the primary vortex street breaks down and is transformed to a secondary vortex.
This secondary expansion occurs when the primary vortex street is in SR.
The vortices of the primary wake are rearranged to a new vortex structure of lower temporal frequency and larger spatial periodicity, accompanying a rapid expansion of $\ell$ and a rapid decrease of $r_K$.
The final approach to the new spatial periodicity $\ell_2$ of the secondary wake is exponential.
As seen with the dotted blue lines in Fig. \ref{fig:transient}(a-e), $\ell(y)$ in this region obeys
\begin{equation}
\ell=\ell_{1}+\left(\ell_{2}-\ell_{1}\right)\left(1-e^{-y/y_{2}}\right),
\label{eq:exponential_expansion}
\end{equation}
where $x_{0}$ is a relaxation length, and our data suggest $y_{2}\sim\ell_{1}$.
The exponential approach is also observed in the initial expansion zone as we stated earlier, suggesting that the underlying physical mechanism of the second instability may not be different from the primary one.
Also, it is inferred that $\ell_1$ of the second expansion is analogous to $D$ of the initial expansion.

In the secondary wake zone, the vortex street is always observed to be CM in our experimental conditions.
The primary wake is apparently shattered and reorganized to form the secondary wake with a larger distance between vortices, as our measurement shows $\ell_{2}\simeq10D$. 
We note that the ratio between the spatial periodicity in the primary and the secondary wake zone, $\ell_{2}/\ell_{1}$, is not an integer.
For instance, with $\Reynolds=$279, 304, 361, 537, and 665, we find $\ell_{2}/\ell_{1}\simeq$2.3, 2.4, 2.4, 2.5, and 2.4, respectively. 
This result is contrary to the prediction from the subharmonic instability theory \citepar{Aref:1981tt, Meiburg:1987gz}. 
Interestingly, $r_K$ gradually increases as in the primary wake zone, suggesting that a sequence of transitions may occur if the flow channel was sufficiently long.

\subsection{Stability criterion}

\begin{figure}
\begin{centering}
\includegraphics[width=8cm]{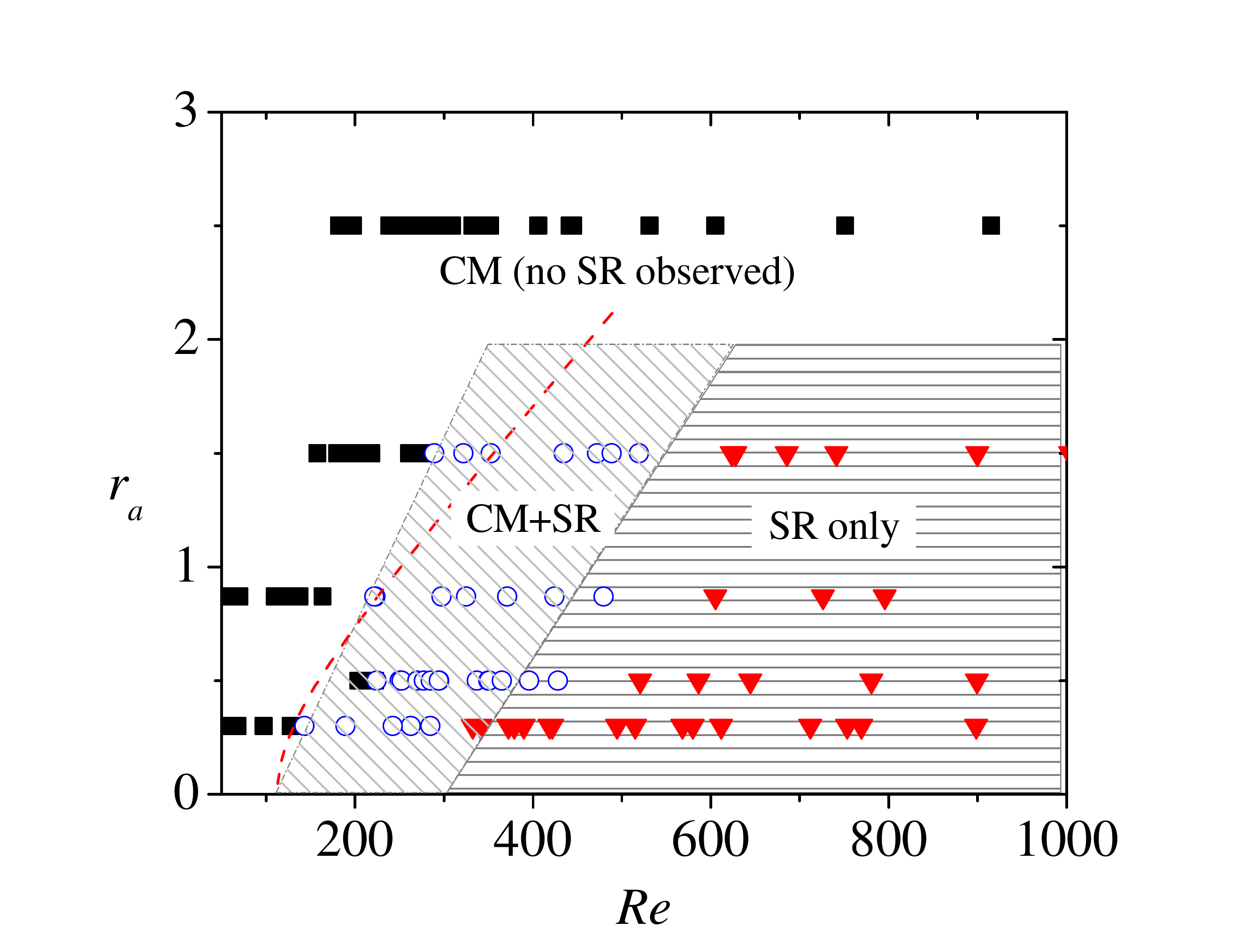}
\par
\end{centering}
\caption{
Spatial structure of vortex streets in the primary wake zone. 
Each point represents the morphological structure of a vortex street created by using a certain $r_a$ and $\Reynolds$; closed squares indicates that the vortex street is in CM, open circles for CM followed by SR, and closed triangles for SR without CM.
The red dashed line is a calculation of Eq. (\ref{eq:onset_secondary_instability}).
\label{fig:phasediagram}}
\end{figure}

In Fig. \ref{fig:phasediagram}, the structures of vortex streets in the primary wake are summarized with respect to $\Reynolds$ and $r_a$.
Our observation shows that there exists a threshold value $\Reynolds_{c2}$, such that if $\Reynolds<\Reynolds_{c2}$, the vortex street is in CM and undifferentiated from the commonly observed vortex streets behind circular objects.
Otherwise, if $\Reynolds>\Reynolds_{c2}$, SR vortex street is observed in the primary wake and the secondary instability occurs downstream.
We name this critical Reynolds number as $\Reynolds_{c2}$ to distinguish it from the onset of the primary wake whose conventional symbol is $\Reynolds_c$.
This notation also reflects the fact that the appearance of SR vortex street coincides with the secondary instability.

We find that $\Reynolds_{c2}$ is an increasing function of $r_a$, i.e. when the rod is more streamlined, higher $\Reynolds$ is required for the onset of the secondary instability.
For instance, when we use the rod of $r_a=0.3$, SR vortex streets are observed if $\Reynolds \gtrsim 130$.
However, when $r_a=1.5$, much higher $\Reynolds\simeq 280$ is required to produce SR vortex streets.
When $r_a=2.5$, the vortex street is always in CM in our range of experiments up to $\Reynolds\simeq 1000$.

Conversely, the observation indicates that the rods with larger $r_a$ is less prone to form SR than the rods with smaller $r_a$ at the same $\Reynolds$ or $D$.
For example, at $\Reynolds=200$, the vortex street is CM when $r_a=1.5$ but SR when $r_a=0.3$.
In the two cases, the height of the triangles differ by factor 5 with the same $D$.
The flow interacts with the boundary for longer time for the rods of higher $r_a$, and the boundary layers are developed thicker for such rods.
As a result, the vortex streets are generated at the different initial condition.

Henceforth, we postulate that the vortex structure in the primary wake zone is determined by the dimensionless ratio $\delta/D$. 
When this ratio is larger than a certain critical value $(\delta/D)_c$, the vortex street is in CM, otherwise the primary wake is partly or entirely SR. 
Our analyses suggest that $(\delta/D)_c \approx 0.4$. 
Using our empirical relation in Eq. (\ref{eq:boundary-layer-thickness}), it is inferred that
\begin{equation}
\Reynolds_{c2}\approx(6/0.4)^{2}\sqrt{r_{a}^{2}+0.25}=225\sqrt{r_{a}^{2}+0.25}.
\label{eq:onset_secondary_instability}
\end{equation}
We mark Eq. (\ref{eq:onset_secondary_instability}) in Fig. \ref{fig:phasediagram} with red dash curve and see that it roughly matches with the experimental measurement.

\section{Conclusion}

We have presented that vortex streets have two morphological configurations, the Conventional Mushroom structure (CM) and the Separated Rows structure (SR).
Most noticeably, SR is characterized by a thin irrotational fluid, which prevents the interaction between the two rows.
On the other hand, CM is characterized by a gradual dissipation of vorticity through the interaction of two vortex rows.

We generated vortex streets in soap film by inserting triangular rods of different aspect ratios.
This technique allowed us to independently adjust the boundary layer thickness $\delta$ and the separation distance $D$ between two boundary layers.
We found that when $\delta/D>0.4$, the vortex street is formed and remains in CM. 
Alternately, when $\delta/D<0.4$, the vortex street is formed in CM and then decays to SR or formed in SR.

Our experiments demonstrate that the spatial structure of the vortex street is determined by the experimental condition.
In previous studies, SR vortex street is perceived as an interim structure that occurs only when a primary vortex street decays and is transformed to the secondary vortex street.
In present study, we find that SR vortex street is formed as the incipient vortex street under certain conditions.
Our observation indicates that the vortex structure is independent of the stability condition of any arrangement.

Lastly, we present that the primary vortex street eventually decays and is transformed to the secondary vortex street.
In downstream, the secondary vortex street evolves in a similar fashion to the primary vortex street. 
We speculate that a sequence of transitions between CM and SR may occur.
In this regard, our study of individual transitions of spatial structures may incur an understanding of the long time evolution of vortex streets, in larger context.

\section*{Acknowledgement}

The experiments was performed at the University of Pittsburgh as a graduate student.
I thank Prof. X. L. Wu for helpful discussions.

\appendix

\section{Observations of Boundary Layers in Soap Films}

A boundary layer is the source of vorticity for downstream vortices, and its detachment from a solid boundary creates the initial condition for the wake formation.
The intensity of the vorticity in the layer and its thickness therefore play an important role for the current discussion.

\begin{figure}
\begin{centering}
\includegraphics[width=8cm]{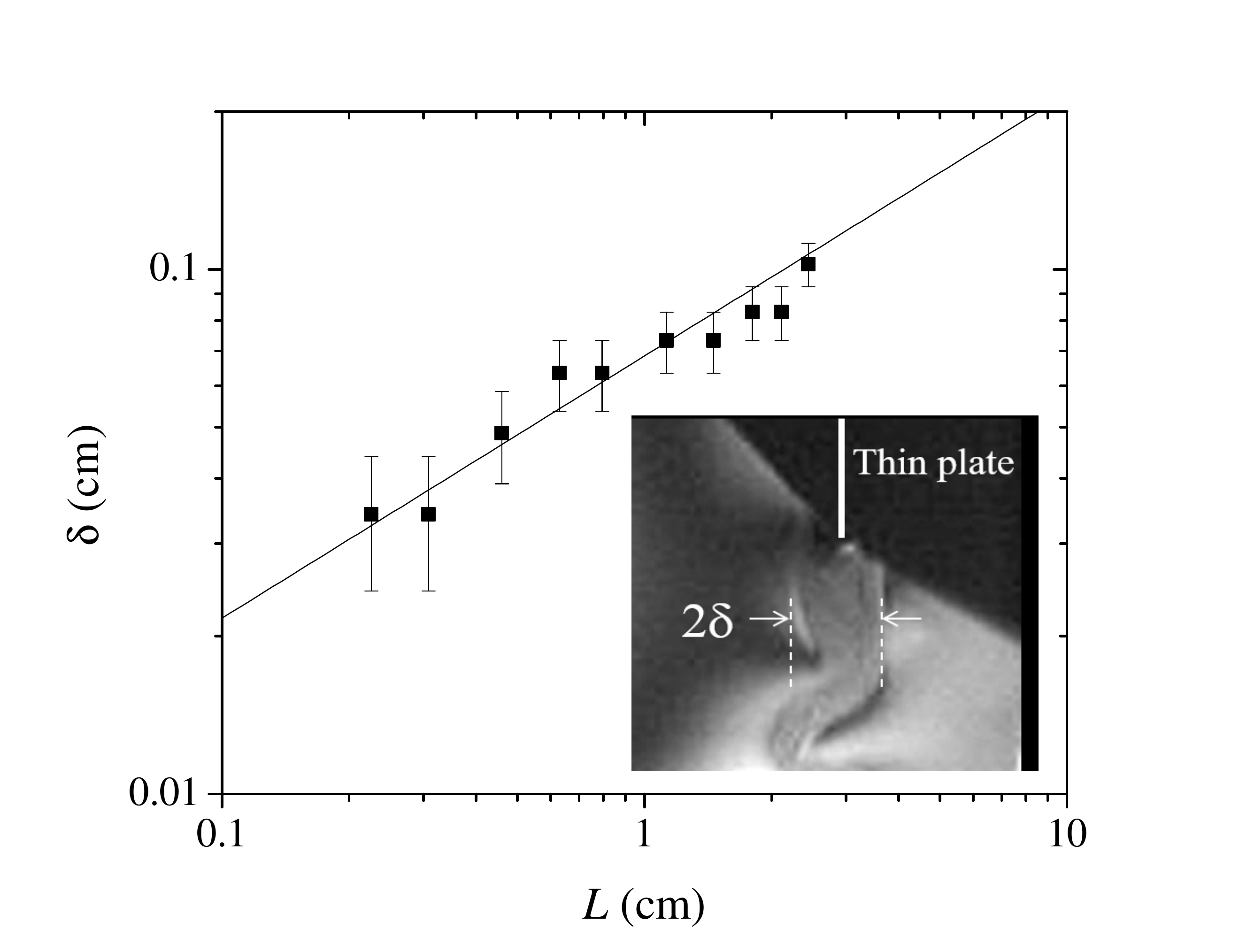}
\par
\end{centering}
\caption{
The measurement of the boundary layer thickness in soap films.
The boundary layer thickness $\delta$ is proportional to $\sqrt{L}$, where $L$ is the longitudinal length of the solid object inserted to the flow.
The solid line is the calculation of Eq. (\ref{eq:estimated_boundarylayer}).
In the inset, an interferogram of boundary layers developed around a thin foril of $L=1.8\,\rm cm$ ($\delta=0.083\,\rm cm$) is shown.
The boundary layers are visualized because the vorticity slightly alters the thickness of the film.
\label{fig:boundarylayer}}
\end{figure}

We measure the boundary layer thickness near a thin plate inserted in flowing soap films using the video imaging technique.
In this investigation, we insert the thin plate perpendicular to the film and parallel to the flow.
The thickness of the thin plate is 25 $\micron$, and the longitudinal length $L$ is varied from 0.2 to 3 cm.
As the soap film flow passes by the thin plate, the boundary layer is developed to a certain thickness $\delta$ depends $L$.
In soap films, the vortical structure in the boundary layer is known to alter the thickness of the film slightly \citepar{Rivera:1998tw, Wu-prl-95} because the medium is slightly compressible \citepar{Kim:2017dn}, and the boundary layer is easily identified in the interferogram, as shown in Fig. \ref{fig:boundarylayer}.

In Fig. \ref{fig:boundarylayer}, our measurement of $\delta$ with respect to $L$ is presented.
The measurement indicates that the boundary layer thickness in our soap film can be approximated to
\begin{equation}
\delta=6\sqrt{\frac{\nu L}{U}}.
\label{eq:estimated_boundarylayer}
\end{equation}
The scaling relation in Eq. (\ref{eq:estimated_boundarylayer}) is consistent to the classical boundary layer theory \citepar{Schlichting}.
We note that the proportionality constant 6 is somewhat greater than the classical value 5.
We believe that the discrepancy is originated from the fact that we use the bulk viscosity value of the soap solution, $\nu=0.013 \,\rm cm^2/s$. 
It is widely known that the surface viscosity of the soap film can greater than the bulk value \citepar{Martin-rsi-95, Vorobieff:1999wn, Prasad:2009jj, Vivek2015}.

For laminar flow passing tapered rods, we approximate the length of the plate to the length of the hypotenuse of the triangle, $L\simeq\sqrt{\left(D/2\right)^{2}+H^{2}}$.
Substituting $L$ in Eq. (\ref{eq:estimated_boundarylayer}), the empirical equation of the boundary layer thickness in Eq. (\ref{eq:boundary-layer-thickness}) is derived.

\bibliography{type2vortex2018}

\end{document}